\documentclass[12pt]{article}

\usepackage{hyperref}

\usepackage{epsf}
\usepackage{amsmath,amssymb}
\usepackage{latexsym}
\usepackage{graphicx,color}
\usepackage{wrapfig}
\usepackage{latexsym}
\usepackage{graphics}

\usepackage{color}
\usepackage{delarray,color}
\usepackage{fancybox}  
\newcommand {\beq}{\begin{align}}
\newcommand {\eeq}{\end{align}}

\newcommand{\be}{\begin{equation}}
\newcommand{\ba}{\begin{align}}
\newcommand{\ea}{\end{align}}
\newcommand{\ee}{\end{equation}}

\newcommand{\beqa}{\begin{align}}
\newcommand{\eeqa}{\end{align}}

\newcommand{\pa}{\partial}

\newcommand{\unit}{\hbox to 3.8pt{\hskip1.3pt \vrule height 7.4pt
    width .4pt \hskip.7pt \vrule height 7.85pt width .4pt \kern-2.4pt
    \hrulefill \kern-3pt \raise 3.7pt\hbox{\char'40}}}

\def\matt[#1,#2,#3,#4]{\left(%
\begin{array}{cc} #1 & #2 \\ #3 & #4 \end{array} \right)}

\newcommand{\ket}[1]{{\left| #1 \right\rangle}}
\newcommand{\bra}[1]{{\left\langle#1\right|}}
\usepackage{tabularx}

\catcode`\@=11
\@addtoreset{equation}{section}

\catcode`@=12
\relax

\begin{document}
%%%%%%%%%%%%%%%%%%%%%%%%%%%%%%%%%%%%%%%%%%%%%%%%%%%%%%%%%%%%%%%%%%%%%%%%
%\baselineskip 0.7cm

\begin{titlepage}

%% Set the number of the title with 0
\setcounter{page}{0}

%% change the footnote symbol
\renewcommand{\thefootnote}{\fnsymbol{footnote}}

\begin{flushright}
%{\tt 
YITP-18-23 \\
%\\}
\end{flushright}

\vskip 1.35cm

\begin{center}
{\Large \bf 
Geometry from Matrices via D-branes
}

\vskip 1.2cm 

{\normalsize
Seiji Terashima\footnote{terasima(at)yukawa.kyoto-u.ac.jp}
}

\vskip 0.8cm

{ \it
Yukawa Institute for Theoretical Physics, Kyoto University, Kyoto 606-8502, Japan
}

\end{center}

\vspace{12mm}

\centerline{{\bf Abstract}}

In this paper, 
we give a map from matrices to a commutative geometry from 
a bound state of a D2-brane and $N$ D0-branes.
For this, tachyons in auxiliary unstable D-brane system
describing the bound state
play crucial roles.
We found the map obtained in this way coincides with 
the recent proposals.
We also consider the map from the geometry
to matrices in a large $N$ limit and
argue that the map is a matrix regularization
of geometry.

\end{titlepage}
\newpage

\tableofcontents
\vskip 1.2cm 

\section{Introduction and Summary}

The D-branes \cite{Pol} in string theory play many important 
roles for applications of string theory to mathematics and field theory.
One of the interesting properties of the D-branes
is that the bound state of two different kinds of BPS D-branes, for example, D0-branes
and D2-branes in type IIA superstring theory,
can be described by two different pictures associated with the D0-branes and the D2-branes.\footnote{
For the bound state with three or more different D-branes,
there are pictures associated with each kinds of D-branes. }
The typical examples of this are the D0-D4 bound state corresponding to the ADHM construction
\cite{Witten:1995gx} \cite{Do2} and the D0-D2 bound state corresponding to the fuzzy sphere
\cite{Myers}.\footnote{This bound state of the D-branes are related to the non-commutative geometry
\cite{CoDoSc, DoHu, SeWi, deWit}.}

There is a systematic way to derive the one picture to the other picture for the bound state, which was
first given in  \cite{Te1}.
This use the tachyon field on an auxiliary unstable D-brane system \cite{Sen} \cite{Senreview}.
In this unstable D-brane system, for example, D2-branes and anti-D2-branes, which we will denote 
as D2-$\overline{\mbox{D2}}$ pairs, %( or D0-$\overline{\mbox{D0}}$ pairs),
%the bound state of the D-branes in 
the D0-brane picture
is represented as a soliton where the tachyon takes 
a non-trivial configuration.\footnote{If the tachyon condenses to a non-zero value,
the D-$\overline{\mbox{D}}$ pairs disappear and if the tachyon field takes a configuration of a topological soliton,
for example the vortex, then the D0-branes remain as a soliton near the zero of the tachyon
\cite{Sen} \cite{Senreview}.
This is called the decent relation and related to the (topological) K-theory \cite{Witten:1998cd}.
}
On the other hand, by diagonalizing the tachyon by a gauge transformation, only the zero mode of the tachyon  
remains after the tachyon condensation \cite{Te1}. %Note that his diagonalization  is just a gauge transformation 
%of D2-$\overline{\mbox{D2}}$ pairs.
This zero mode in the D2-$\overline{\mbox{D2}}$ pairs
corresponds to a D2-brane. 
Thus, we find the D2-brane picture from the D0-brane picture
through the D2-$\overline{\mbox{D2}}$ pairs.
The role of the D2-brane and the D0-brane can be switched and 
we can have the D0-brane picture from the D2-brane picture.\footnote{
For this, we need the higher dimensional ``soliton'' on the D0-$\overline{\mbox{D0}}$ pairs,
which was found in \cite{Te0, AST1, AST2}, and this construction of the higher dimensional branes 
is called the ascent relation, which is related to the (analytic) K-homology.
}

This change of the pictures or the descriptions of the bound state 
can be done in the boundary superstring field theory \cite{KMM2, KrLa, TaTeUe} or 
the (off-shell) boundary state.
Indeed, in \cite{AST3},
we showed that the decent relations and the ascent relations exactly, i.e. 
the equivalences between the boundary state of  the unstable D$p$-brane system
and the one of the BPS D$q$-brane.
Using \cite{AST3}, 
the diagonalization of the tachyon gives the equivalence between 
the boundary state of the D0-branes 
and the one of the D2-brane explicitly, where all the string scale $l_s$ corrections are 
included.
This method was applied to the Nahm equation \cite{HaTe3}, ADHM construction
of instantons \cite{HaTe4}, the supertubes \cite{Te2} 
and Nahm transformation \cite{Te3}.

As a special application of this method,
a map from a geometry to matrices is obtained \cite{Ell, Te1}.
Indeed, by using the method, we can show that 
the D$(2p)$-brane wrapping $2p$-dimensional sub-manifold
with a flux which is a two form, is equivalent to
the D0-branes with matrices which represent the positions.
However, in general, the (usual) geometry does not describe the physics well 
because the system is described by a fuzzy geometry from the D0-brane picture
and the boundary state of the compact D2-brane used in \cite{Ell, Te1} 
may not need to correspond to the D2-brane described by the DBI action on the geometry
even if the curvature of it is small.\footnote{
In particular, the massive string modes may be excited
because most of them are not coupled to the RR-fields \cite{HST}.
}
A D-brane probe or closed string scattering amplitudes will see the fuzziness, for example \cite{Ha}.
It has been expected and checked that 
fuzziness is related to the finiteness of $N$ and 
the geometrical D2-brane is
obtained 
in a suitable large $N$ limit, where $N$ is the D0-brane charge.

In this paper, we apply the method 
to the bound state of the D2-brane and the D0-branes
via the D2-$\overline{\mbox{D2}}$ pairs.
This gives the inverse of \cite{Te1}, i.e. a geometry from the matrices.
For this, we need a special class of large $N$ matrices
which correspond to a commutative geometries.
The map obtained in this way coincides with 
the recent proposals \cite{Ber, Ishiki, Stei}.
We also consider the map from the geometry
to matrices using the method in the large $N$ limit.
The algebra of the functions on a smooth manifold 
can be approximated by matrices, which is called 
the matrix regularization of geometry, see, for example, \cite{mr}. We argue the map is a matrix regularization.

The organization of this paper is the following.
In section 2, for suitable matrices,
we derive the map from the matrices to a geometry
using the tachyon condensation in
the large $N$ limit.
We also discuss the large $N$ limit of the map 
from the geometry to matrices in section 3.

\section{Geometry from Matrices via D-brane}

\subsection{Non-commutative plane}

In this subsection, we will consider 
a simplest example, corresponding to the non-commutative plane, of the mapping from
the matrices to the geometry, which was shown in the appendix in \cite{Te2}.
In a D0-brane picture, 
the system we consider is the large $N$ limit of $N$ D0-branes with matrix coordinate $X^i$:
\begin{align}
X^1=\hat{x}^1, \,\,\, X^2=\hat{x}^2, \,\,\, \mbox{Others}=0,
\label{ncp}
\end{align}
where $X^i$ is the matrix valued coordinates of D0-branes,
\begin{align}
[\hat{x}^1, \hat{x}^2]={i\over B}  \unit. %{\mathbf 1}.
\end{align}
Here, $B$ is a constant number, $\unit$ is the identity operator\footnote{
We will often omit the identity operator below.} 
and an operator $\hat{x}^i$ is regarded 
as an infinite dimensional matrix in the large $N$ limit.\footnote{
This is called non-commutative plane.}
It is well-known that this system is the bound state of 
the $N$ D0-branes and a D2-brane. 
In the D2-brane picture, this will be represented by the D2-brane 
with a background magnetic flux.
These two different pictures are explicitly shown to be equivalent
using auxiliary unstable D-brane systems in \cite{Te1}.
Below we will apply this to (\ref{ncp}) to obtain the D2-brane picture
which is described by a geometry, instead of the matrices.

First, we note that a D0-brane at a point $X^1=a^1, X^2=a^2$
is equivalent to a pair of a D2-brane and an anti-D2-brane %, which we will denote 
%as D2-$\overline{\mbox{D2}}$ pair,
with the following non-trivial tachyon:
\begin{align}
\matt[0,T,T^\dagger,0]=u \left(
\sigma_1 (x^1-a^1)+\sigma_2 (x^2-a^2)
\right),
\end{align}
where the D2-$\overline{\mbox{D2}}$ pair extends in $x^1,x^2$ plane
(except the time direction) and $\sigma_i$ is the Pauli matrix.
This configuration is equivalent to the D0-brane 
in the $u \rightarrow \infty$ limit as the boundary state which 
includes $\alpha'$ corrections \cite{AST3}.
Here, the $u \rightarrow \infty$ limit implies that
the D2-$\overline{\mbox{D2}}$ pair is annihilated and 
the D0-brane remains as a soliton (voltex) on the pair.

Using this, for the infinitely many D0-branes with (\ref{ncp}),
we obtain the infinitely many D2-$\overline{\mbox{D2}}$ pairs
with the tachyons:
\begin{align}
\matt[0,T,T^\dagger,0]=u \left(
\sigma_1 (x^1-\hat{x}^1)+\sigma_2 (x^2-\hat{x}^2)
\right), 
\end{align}
where $T$ is an $N \times N$ matrix
and the gauge fields on the D2-$\overline{\mbox{D2}}$ pairs are zero.

In order to obtain the D2-brane picture (which is geometrical in some sense),
we need to diagonalize the tachyon
(for the Chan-Paton indices) and  
consider only the zero modes of the tachyon \cite{Te1}
because non-zero modes will disappear in
the $u \rightarrow \infty$ limit.\footnote{
Note that the D-$\overline{\mbox{D}}$ pair completely disappears for $T=\infty$ 
as first Sen conjectured \cite{Sen}.}
Thus, we need to solve the following zero-mode equation:
\begin{align}
0=\matt[0,T,T^\dagger,0] 
\left( \begin{array}{c} \psi_1(z)  \\ \psi_2(z)  \end{array} \right)
\sim \matt[0,\sqrt{B/2} \bar{z}-\hat{a}^\dagger, \sqrt{B/2} z - \hat{a},0] 
\left( \begin{array}{c} \psi_1(z)  \\ \psi_2(z)  \end{array} \right),
\end{align}
where $z=x^1+i x^2$ and $\hat{a}=\sqrt{B/2} (\hat{x}^1+i \hat{x}^2)$.
Note that here $z$ is a parameter, not an operator.
Assuming $B>0$, which implies $[\hat{a}, \hat{a}^\dagger]=\unit$,
we easily find that there is unique zero mode:
\begin{align}
\psi_1(z) = \ket{\sqrt{B/2} z}, \,\,\, \psi_2(z) =0,
\end{align}
where $\hat{a} \ket{z}=z \ket{z}$ 
is the coherent state.
This normalized state parametrized by $z$ corresponding to
the D2-brane extending in the $x^1,x^2$ plane 
which remains after the tachyon condensation on the D2-$\overline{\mbox{D2}}$ pairs.

The gauge field on the D2-brane is given by 
$A_i =i \psi_1^\dagger {\partial \over \partial x^i} \psi_1$ \cite{AST3}.
In our case, using 
$\ket{z}= e^{-|z|^2/2 +z \hat{a}^\dagger} \ket{0}$,
we find
\begin{align}
A_z=i \bra{\sqrt{B/2} z} \partial_z \ket{\sqrt{B/2} z}=i {\bar{z} \over 4} B,
\label{gf1}
\end{align}
where $A_z=\frac12 (A_1-iA_2)$ and $\partial_z=\frac12 (\partial_1-i \partial_2)$.
The field strength is $F_{z\bar{z}} =\partial_z A_{\bar{z}} - \partial_{\bar{z}} A_{z}=-i B/2$, or
in the original coordinates,
\begin{align}
F_{12}=2 {\partial z \over \partial x^1 } {\partial \bar{z} \over \partial x^2 } F_{z \bar{z}}=-B,
\end{align}
which is proportional to the D0-brane density through the Chern-Simons couplings
on the D2-brane.
Note that in a large D0-brane charge density limit. i.e. a large $B$ limit,
where the bound state will be described mainly by the D0-branes,
the matrix valued positions of the D0-branes nearly commute each other,
$[X^1,X^2] \sim 0$. Note that even in this limit sign of the flux $B$ is important 
to determine the D2-brane charge.

We obtained the D2-brane picture (or geometrical picture)
of the D0-branes with non-commutative coordinates.
For the later generalization to the non-flat case,
we will introduce (inifinitely many) non-BPS D3-brane
extending in $x^1,x^2,x^3$ plane to describe the D2-D0 bound state.
This type of construction was first used in \cite{HaTe3} for the derivation of the Nahm transformation.
First, the D0-branes with (\ref{ncp}) is equivalent to
the non-BPS D3-brane with the following tachyon (after taking the $u' \rightarrow \infty$ limit) \cite{AST3}:
\begin{align}
T_{D3}= u' \sum_{i=1}^3 \sigma_i  (x^i \unit_{N \times N} -X^i),
\label{td3}
\end{align}
where %$i=1,2,3$ and 
$X^3=\zeta^3$. 
Although $\zeta^3=0$ for (\ref{ncp}),  
here we generalized it so that the position of the D0-branes is at $x^3=\zeta^3$
for later convenience.
The non-BPS D3-brane with the tachyon $T_{D3}$ can be obtained from the 
the $2 k N$ D2-$\overline{\mbox{D2}}$ pairs extending in $x^1,x^2$ plane
with $X^3=\zeta^3$ and the tachyon
\begin{align}
T' & =
%u(i \hat{p}^3 \otimes \unit_{2 \times 2} \otimes  \unit_{N \times N}
%+ \unit_{k \times k} \otimes T_{D3}) 
u \left( { \partial \over \partial x^3} 
+  T_{D3}\right)
%\nonumber \\
=
u \left( { \partial \over \partial x^3} 
+  
u' \sum_{i=1}^3 \sigma_i  (x^i \unit_{N \times N} -X^i)
\right)
\label{t3}
\end{align}
where $x^3$ and $ { \partial \over \partial x^3} $ are regarded as $k \times k$ matrices
in the large $k$ limit.
More explicitly, 
\begin{align}
T' & =
u \left( { \partial \over \partial x^3} +u' \sigma^3 (x^3-\zeta^3)\right) +
u u' \left(
\sigma_1 (x^1-\hat{x}^1)+\sigma_2 (x^2-\hat{x}^2)
\right).
\end{align}
Then, we can see that the only
the normalizable zero-mode of the tachyon, $T' \psi=0$,
is 
\begin{align}
\psi=\left( {u'  \over \pi} \right)^{\frac14} e^{-u' (x^3-\zeta^3)^2/2} 
\left( \begin{array}{c} 1  \\ 0  \end{array} \right)
\ket{\sqrt{B/2} z}
\rightarrow  \sqrt{\delta(x^3-\zeta^3)} \left( \begin{array}{c} 1  \\ 0  \end{array} \right)
\ket{\sqrt{B/2} z},
\end{align}
where we took the $u' \rightarrow \infty$ limit.
(There is no normalizable zero-modes for $(T')^\dagger \psi=0$).
This is localized on $x^3=\zeta^3$  because of the $u' \rightarrow \infty$ limit
and represents the D2-brane.
The gauge field on the D2-brane is computed by
\begin{align}
A_i=i \int d x^3 \psi^\dagger  { \partial \over \partial x^i} \psi,
\end{align}
where $i=1,2$, which coincides with (\ref{gf1}).
The scalar field $\Phi^3$ on the D2-brane which represents the position in $x^3$ direction
is given by
\begin{align}
\Phi^3=i \int d x^3 \psi^\dagger x^3 \psi=\zeta^3,
\end{align}
as expected.

\subsection{Curved case}

Let us consider $N$ D0-branes with the general $X^i$ $(i=1,2,3)$.
Then, as in the previous subsection, 
the D2-brane picture can be obtained by
considering the non BPS D3-branes and the corresponding 
D2-$\overline{\mbox{D2}}$ pairs extending in $x^1,x^2$ plane
with the tachyon $T'$ given by (\ref{t3}).
Then, denoting the number of  the zero-modes of the tachyon $T'$ 
as $N_2$, the system is equivalent to the $N_2$ D2-branes 
with the scalar $\Phi^3$ and the gauge fields $A_i$ on them,
which are $N_2 \times N_2$ matrices and given by
\begin{align}
(A_i)_{\rho \rho'} &=i \int d x^3  (\psi_\rho)^\dagger { \partial \over \partial x^i} (\psi_{\rho'}),
\nonumber \\
(\Phi^3)_{\rho \rho'}&=i \int d x^3 (\psi_\rho)^\dagger x^3 (\psi_{\rho'}),
\end{align}
where $\psi_\rho$ ($\rho=1,2,\cdots, N_2$) are the normalized zero modes
and $i=1,2$.
This procedure gives the exact mapping from the D0-brane boundary state
to the D2-brane boundary state.

However, 
the geometrical D2-brane picture 
is not a good description of
the bound state of D2-branes and D0-branes, in general.
For example, the fuzzy sphere configuration
with $N=2$ D0-branes (or a few D0-branes) is not 
well described by a smooth D2-brane.
The fuzziness of the finite $N$ system
can be understood, for example, from the coupling 
to the RR fields \cite{Ha}.\footnote{
Note that there will be a D2-brane picture even for the fuzziness is not small.
Here, a good description means that
the DBI action of the D2-brane is a good approximation of the system
and other corrections including the higher derivative terms are small.}
Imposing some special conditions on the scalars of the D0-branes, 
we will expect the system corresponds to a smooth D2-brane
and there should be a map from the D0-branes to the D2-brane, i.e.
the matrices to the geometry.
These conditions includes a large $N$ limit which have not been studied
in the previous studies in the tachyon condensation point of view.

Below, we will consider the cases the system is 
well described by a smooth geometry (i.e. a D2-brane) and give this explicit map.
For this, the system should be locally approximated by
the system considered in the previous section, i.e. the non-commutative plane,
up to spacetime rotations and translations.
To suppress the fuzziness, we further need to require that $B$ is large,
which means the large D0-brane density.\footnote{
More precisely, we require $B ( l_s)^2 \gg  1$ and $l_{D2} \gg l_s$
where $l_{D2}$ is a typical scale of the geometry given by the D2-brane,
for example. the size of the sphere for the fuzzy sphere.
Note that $N\sim B (l_{D2})^2$ for a compact manifold.
}
These mean that by a rotation, $R$, and a translation $d$,
the scalars satisfy
\begin{align}
[{X'}^1, {X'}^2] \approx {i\over B}  \unit, \,\,\, {X'}^3 \approx 0,
\label{com1}
\end{align}
where ${x'}^i=\sum_j^3 R^i_{\,\, j} {x}^j +d^i$,
\begin{align}
{X'}^i=\sum_j^3 R^i_{\,\, j} {X}^j +d^i,
\label{rot}
\end{align}
and the neglected terms become small in the large $B$ limit.
Here, these relations are satisfied for a subset of the Hilbert space of the states,
which corresponding to a local region of a point in spacetime.

Thus, for such D0-branes, 
we have an approximate description by a D2-brane.
The geometry of the D2-brane is given by a collections of the points where
${X'}^1={X'}^2={X'}^3=0$, which is equivalent to the points in 
the space parametrized by $\{ x^1,x^2,x^3 \}$ where 
$T_{D3}$ in (\ref{td3}) has zero modes, i.e. 
$T_{D3} \psi=0$.
Note that in the rotated coordinate,
\begin{align}
T_{D3}= u' \sum_{i=1}^3 {\sigma'}_i  ({x'}^i \unit_{N \times N} -{X'}^i),
\label{td3a}
\end{align}
where ${\sigma'}^i=\sum_j^3 R^i_{\,\, j} {\sigma}^j $,
The gauge field strength at the point is given by $F_{1' 2'}=-B$
where $B$ in (\ref{com1}) varies on the D2-brane
and $F_{1' 2'}$ is computed in the $x'$-coordinates.
Because $T_{D3}$ corresponds to the Hamiltonian (or the Dirac operator)
in \cite{Ber, Ishiki, Stei},
the geometry and the 2-form are same as those given in  \cite{Ber, Ishiki, Stei}.

In summary, the map from the $N \times N$ matrices $X^i$ to a smooth surface is given 
by the followings. First, let us define $2 N \times 2N$ matrix $ D=\sum_{i=1}^3 {\sigma}_i  ({x}^i \unit_{N \times N} -{X}^i)$ and consider a linear equation $D \psi=0$ where $x^i$ are regarded as parameters
and $\psi$ is a $2N$ vector which depends on the parameter $x^i$.
Then, the surface is given by the collection of the points $x^i$ for which 
the equation has non-zero solutions.

Below, we will consider the system more explicitly using the boundary state.
In order to do it, we need to fix the plane on which the 
D2-$\overline{\mbox{D2}}$ pairs extending because 
the boundary state usually is defined for the flat D-branes.\footnote{
The D-branes on curved manifold can be considered, for example as in \cite{Te1},
however, there are many ambiguities for the definitions.
The on-shell condition for the boundary state, i.e. the conformal condition,
is also difficult to be studied.
}
We take this as $x^1,x^2$ plane and
parametrize the sub-manifold (on  which the D2-brane exist)
defined by the points where zero modes of $T_{D3}$ exist and we will denote this sub-manifold as
${x}^3={\zeta}^3(x^1,x^2)$.
For a point in the sub-manifold, 
the orthogonal coordinate system ${x'}^i$ is associated
such that ${x'}^1={x'}^2={x'}^3$ is on the sub-manifold and
${x'}^1,{x'}^2$ (and ${x'}^3$) are tangent (and perpendicular) to the sub-manifold at the point, respectively.
Then the tachyon on the D2-$\overline{\mbox{D2}}$ pairs 
is
\begin{align}
T' & =
u \left( { \partial \over \partial x^3} 
+  
u' \sum_{i=1}^3 {\sigma'}_i  ({x'}^i \unit_{N \times N} -{X'}^i)
\right),
\label{t3a}
\end{align}
with (\ref{com1}) 
where $B$ is considered as functions of $x^1,x^2$.
This is approximated as
$T' \sim
u \left( { \partial \over \partial x^3} 
+  u' {\sigma'}_3 {x'}^3(x^1,x^2) +
u' \sum_{i=1}^2 {\sigma'}_i  ({x'}^i \unit_{N \times N} -\hat{x}^i)
\right),
$
at a point in ${\{ x^1,x^2 \}}$ plane.
Using the result for the flat case, the zero mode of this is easily obtained 
by the rotation at each point as
\begin{align}
\psi=
\sqrt{\delta({x}^3-\zeta^3(x^1,x^2))} \Gamma(R(x^1,x^2)) \left( \begin{array}{c} 1  \\ 0  \end{array} \right)
\ket{\sqrt{B(x^1,x^2)/2} \, z({x'}^1,{x'}^2)},
\end{align}
where $z({x'}^1,{x'}^2)={x'}^1+i {x'}^2 $
and $ \Gamma(R(x^1,x^2))$ is the rotation matrix for the spinor, which satisfies
$ {\Gamma(R(x^1,x^2))}^{\dagger} {\sigma'}_i  \Gamma(R(x^1,x^2)) = {\sigma}_i$.\footnote{
The contribution from ${ \partial \over \partial x^3} $ in $T'$ acting on 
$\ket{\sqrt{B(x^1,x^2)/2} \, z({x'}^1,{x'}^2)}$ is negligible  because it is $u'$-independent.
}
Here, the delta-function $\sqrt{\delta({x}^3-\zeta^3(x^1,x^2))}$ appeared from
the $\left( {u'  \over \pi } \right)^{\frac14} \exp(-{ u'   ({x'}^3)^2  \over (R^3_{\,\, 3})^2 })$ in the $u' \rightarrow \infty$
limit because $\delta({x'}^3)$ is proportional to $\delta({x}^3-\zeta^3(x^1,x^2))$.

{}From this zero-mode, we can compute the scalar which represents the 
$x^3$ position as
\begin{align}
\Phi^3 =i \int d x^3 (\psi)^\dagger x^3 \psi =\zeta^3(x^1,x^2),
\end{align}
which means the D2-brane indeed wrapping the sub-manifold.
For the gauge field on the D2-brane, we have
\begin{align}
A_i=i \int d x^3 \psi^\dagger  { \partial \over \partial x^i} \psi, \,\,\, (i=1,2),
\end{align}
where $ { \partial \over \partial x^i}$ should be taken with fixing $x^3$.
For the large $B$ approximation, which includes $l_{D_2} \partial_i B \ll B$,
the computation reduces to the one in the previous section and we find
\begin{align}
A_i=i  \int d x^3  { \partial z \over \partial x^i} \psi^\dagger  { \partial \over \partial z} \psi
+i  \int d x^3  { \partial \bar{z} \over \partial x^i} \psi^\dagger  { \partial \over \partial \bar{z}} \psi
= \left. { \partial z \over \partial x^i} \right|_{x^3=\zeta^3} A_z 
+\left.  { \partial \bar{z} \over \partial x^i}\right|_{x^3=\zeta^3}  A_{\bar{z}},
\end{align}
where 
\begin{align}
A_z \approx  {1\over 4}  \left.  \bar{z}  B(x^1,x^2)\right|_{x^3=\zeta^3}.
\end{align}
Thus, this describes the D2-brane on the sub-manifold with
the gauge field strength $F_{z\bar{z}}=-B$ locally as expected.
Note that for D2-brane on a compact sub-manifold, 
this flat space boundary state description becomes singular at some points,
where the D2-brane is perpendicular to the $x^1,x^2$ plane.
Actually, for a compact manifold, the scalars describing the position 
should be highly non-trivial.
Here, the description of a D2-brane will be valid because it is only
an approximation in the large $N$ limit.

\subsection{General surfaces}

In the previous subsection, 
we considered the two dimensional surface in three dimension.
It is trivial to generalize this to a surface in higher dimensions
by introducing the non-BPS D$(2n+1)$-branes instead of the non-BPS D3-branes.
Furthermore, the  bound state of D0-branes and D4-brane can have
a smooth geometry description.\footnote{
We can easily generalize this to a D$(2n)$-brane of course.
For a odd dimensional sub-manifold, non-BPS D-brane might be play a role.
For example, a D0-brane at the origin is equivalent to a non-BPS D1-brane extending 
in $x^1$ direction with $T=u x^1$. However, this geometry is not unique, thus it is unphysical
because we can rotate the non-BPS brane or deform the tachyon far from the origin.
If the D0-branes are equivalent to a non-BPS D-brane with $T=0$, we can associate 
a odd dimensional geometry although it is highly difficult.
 }
For this case, instead of (\ref{com1}), we need
\begin{align}
[{X'}^1, {X'}^2] \approx {i\over B^{(1)}}  \unit, \,\,\, [{X'}^3, {X'}^4] \approx {i\over B^{(2)}}  \unit, \,\,\, 
\mbox{Othres} \approx 0.
\label{com2}
\end{align}
with which the system is described by a D4-brane on
a four dimensional sub-manifold.

\section{Matrices from Geometry }
\label{MG}

In this section, we will consider the  
inverse mapping of the previous section.
We have seen how the geometry can be obtained from the matrices
for the certain class of the bound states of D0-D2-branes through the tachyon condensation.
Because we can map the D2-brane picture to the D0-brane picture,
the inverse, i.e. the matrices from the geometry, is also possible.
Indeed, this was done in \cite{Te1} by the tachyon condensation.\footnote{ 
Without the interpretation of the tachyon condensation, this mapping was derived in \cite{Ell}.}
In \cite{Te1}, the D2-brane on a curved manifold is represented 
in the infinitely many D0-$\overline{\mbox{D0}}$ pairs where 
the tachyon becomes the Dirac operator on the manifold including the gauge field of the D2-brane.
Then, in the basis of the Chan-Paton index where the tachyon
is diagonalized, the zero-modes of the tachyon remains.
These zero-modes represent the D0-branes, and we will denote the number of them as $N$.
The matrix valued scalars representing the positions of the D0-branes
can be computed from the zero-modes.
Thus, what we will consider is the large $N$ limit, i.e. the large flux limit, of this procedure.\footnote{
The following discussions might be rigorously stated as the
$Spin^{ C}$ quantization of manifold. % \cite{}. 
Here, we will give some intuitive and 
stringy explanations.}

We will start from the D$(2n)$-brane in a spacetime ${\mathbf R}^D$. 
The geometric data of it are the followings:
a $2n$-dimensional closed $Spin^{ C}$ manifold ${\mathcal M}_{2n}$,
an embedding $\varphi: {\mathcal M}_{2n} \rightarrow {\mathbf R}^D$
which is given by $x^I (\sigma^\mu)$ where $I=1,2, \cdots,D$ and $\mu=1,2, \cdots,2n$
and a non-degenerate closed 2-form $\omega$ on ${\mathcal M}_{2n}$.
Here, $\omega$ is related to the field strength of the D$(2n)$-brane,
which should be quantized. 
Because $N$ is the D0-charge, 
the field strength increases infinitely
in the large $N$ limit. 
We will consider a sequence of the D$(2n)$-branes which are parametrized by
a flux $N$ and denote the $U(1)$ connection on ${\mathcal M}_{2n}$ as $A^{(N)}$
which satisfies ${1 \over n!} \int_{{\mathcal M}_{2n}} (d A^{(N)} )^n =N$.\footnote{
More precisely, we require $\int_{{\mathcal C}_{2}} (d A^{(N)} ) =N_{{\mathcal C}_{2}} \in {\mathbf Z}$ 
for any non-trivial two-dimensional surface ${\mathcal C}_{2}$ in ${\mathcal M}_{2n}$ and take the large $N_{{\mathcal C}_{2}}$ limit. 
}
In the large $N$ limit, we can approximate any $\omega$ (which is $N$ independent) by the gauge field, such that
$\left( \alpha/N \right)^{1 \over n} d A^{(N)} \rightarrow   \omega $ in the $N \rightarrow \infty$ limit
where $\alpha= %{1 \over N} 
{1 \over n!} \int_{{\mathcal M}_{2n}} \omega^n $
and $N$ will not need to take every natural number.
We will take $\alpha=1$ by a rescaling of $\omega$.

With this and the induced metric 
$g_{\mu \nu}=\partial_\mu x^I(\sigma) \partial_\nu x^I(\sigma)$, we can define the Dirac operator 
\begin{align}
D^{(N)} \equiv i \gamma^\mu (\pa_\mu-i A_\mu^{(N)}),
\end{align}
where a vierbein which is consistent with the induced metric 
was chosen, and consider 
the eigen modes of the Dirac operator
such that
\begin{align}
D^{(N)} \ket{i,a}^{(N)} = E_i \ket{i,a}^{(N)}, \,\,\,\,   {}^{(N)} \left\langle i,a \right.  \ket{j,b}^{(N)}=\delta_{ij}\delta_{ab}.
\end{align}
{}From of the index theorem, we find there are $N$  
zero-modes of the Dirac operator, $\ket{0,a}^{(N)}$, where $a=1,2, \cdots,N$.
With these, we can define $N \times N$ matrices, $X^I_{(N)}$, as
\begin{align}
(X^I_{(N)})^a_{\,\, b}   \equiv  {}^{(N)} \left\langle 0,a \right| x^I(\sigma^\mu) \ket{0,b}^{(N)},
\label{m1}
\end{align}
where $I=1,2, \cdots,D$ and $a=1,2, \cdots,N$.
These matrices give the non-commutative or fuzzy geometry
and we expect that these give a matrix regularization of geometry \cite{mr}
although spacetime ${\mathbf R}^D$ and the embedding $x^I(\sigma^\mu)$ may not be necessary
for the interpretation of the matrix regularization of ${\mathcal M}_{2n}$
in the large $N$ limit.

More generally, we can define a map $W: \, C^0({\mathcal M}_{2n}) \rightarrow M_{N}(\mathbf C) $,
where $M_{N}(\mathbf C) $ is the set of the complex valued $N \times N$ matrices. 
Let us consider the continuous function $f(\sigma)$ on ${\mathcal M}_{2n}$.
Denoting $W(f) =\hat{f}$, we define
\begin{align}
(\hat{f} )^a_{\,\, b}    \equiv  {}^{(N)} \left\langle 0,a \right| f(\sigma) \ket{0,b}^{(N)}.
\label{W0} 
\end{align}
For the continuous function $\tilde{f}(x^I)$ on the spacetime ${\mathbf R}^D$,
we can define the continuous function on ${\mathcal M}_{2n}$, $f \equiv \tilde{f} \circ \varphi$, 
i.e. $f(\sigma^\mu)=\tilde{f}(\varphi(\sigma^\mu))$, and then we can map it to a matrix by (\ref{W0}).\footnote{
Another definition of $\tilde{W}: \, C^0({\mathbf R}^D) \rightarrow M_{N}(\mathbf C) $
might be
$
\hat{f}= \int d p^I  f(p^I) e^{i p^I X^I}, 
$
where $f(p^I)$ is the Fourier decompision of $f(x^I)$. i.e. $f(x^I)= \int d p^I  f(p^I) e^{i p^I x^I}$,
and $X^I$ is the matrix defined by (\ref{m1}).
This means that $\hat{f}=f(X^I)$ with the ``symmetric'' ordering which is the Weyl ordering if the $X^I$ satisfy the 
canonical commutation relation.
This might be related to \cite{Shimada}.
We can take another appropriate ordering instead of the this ordering. 
%Here, we consider only some class of functions for which 
%the definition is well defined.  
Note that  $\tilde{f}_1 =  \tilde{f}_2$ is possible for $f_1 \neq f_2$ where
$\tilde{f}_1 =f_1 \circ \varphi,$ and $\tilde{f}_2 =f_2 \circ \varphi$.
Thus, there are ambiguities for a map from $C^0({\mathcal M}_{2n})$ to $C^0({\mathbf R}^D)$,
and we choose one of them.
If there are no self-intersections for the embedding of ${\mathcal M}_{2n}$,
any function on ${\mathcal M}_{2n}$ can be constructed from
$\{x^I(\sigma), I=1,2,\cdots,D \}$.
Then, one choice of a map from $C^0({\mathcal M}_{2n})$ to $C^0({\mathbf R}^D)$
is given by the identification of $x^I(\sigma)$ as $x^I$.
For example, for two-sphere, $x^1(\sigma)=\sin \sigma_1 \sin \sigma_2, x^2(\sigma)=\sin \sigma_1 \cos \sigma_2,x^3(\sigma)=\cos \sigma_1$, with which we can construct the spherical harmonics,
and $x^1=r \sin \sigma_1 \sin \sigma_2, x^2=r \sin \sigma_1 \cos \sigma_2,x^3=r \cos \sigma_1$.
}
Below, we will argue that the matrices satisfy the following properties
for $f_i \in C^0({\mathcal M}_{2n})$:
If $f_1 f_2 =f_3$,
then 
\begin{align}
\hat{f}_1^{(N)} \hat{f}_2^{(N)} -\hat{f}_3^{(N)} \rightarrow 0, 
\label{c1}
\end{align}
for $N \rightarrow \infty$, where %$\tilde{f}_i= f_i \circ \varphi$ and 
$\hat{f}_i= W(f_i )$.
We will also argue that the following stronger statement holds for $N \rightarrow \infty$:
\begin{align}
N^{1 \over n} \left( [ \hat{f}_1^{(N)},  \hat{f}_2^{(N)} ] -
W \left( i \theta^{\mu \nu} {\partial \over \partial \sigma^\mu}  f_1^{(N)} (\sigma) {\partial \over \partial {\sigma}^\nu}   f_2^{(N)}  (\sigma) \right)
\right)
\rightarrow 0, 
\label{c2}
\end{align}
where
%$(f_1^{(N)} * f_2^{(N)}) (\sigma) \equiv 
%i \theta^{\mu \nu} {\partial \over \partial \sigma^\mu}  {\partial \over \partial {\sigma'}^\nu} (f_1^{(N)} (\sigma) f_2^{(N)} (\sigma') )|_{\sigma'=\sigma}$
%and 
$\theta^{\mu \nu}$ is anti-symmetric tensor which 
is the inverse of $d A^{(N)}$, i.e. 
$\theta^{\mu \nu} \, N^{1 \over n} \omega_{\nu \mu'}=\delta^\mu_{\,\, \mu'}$.
Here, the algebra of the functions defines the geometry and 
(\ref{c1}) means that the matrix approximation of the algebra indeed equivalent
to the original algebra in the large $N$ limit.
Furthermore, (\ref{c1}) means that the commutator is equivalent to
the Poisson bracket defined by the $\theta^{\mu \nu}$ in the large $N$ limit.

Below, we will concentrate on $n=1$ case, i.e. the  D2-brane case because
generalizations are straightforward.
Because ${1 \over N} d A^{(N)} \rightarrow   \omega $,
we can take a constant $c$ such that
$| d A^{(N)}| > c N $ on ${\mathcal M}_{2}$, 
where $c$ is an $N$ independent constant on ${\mathcal M}_{2}$
and $| d A^{(N)}| =\sqrt{g^{\mu \mu'}g^{\nu \nu'}(d A^{(N)})_{\mu \nu}  (d A^{(N)})_{\mu' \nu'}  }$. 
Let us consider a region at a point in ${\mathcal M}_{2}$ and the size of the region, $l$, satisfies
\begin{align}
l_B \ll l \ll l_{D2},
\label{size}
\end{align}
where
$l_{D2} \sim \sqrt{\det (g)}$ is a typical length scale of 
the embedding and $l_B =\mbox{min} \{  1/\sqrt{| d A^{(N)} |} \}$ is a length scale of the flux.
Here, $\mbox{min} $ means the minimum value for all points in ${\mathcal M}_{2}$
and we expect $l_B \sim {l_{D2} \over \sqrt{N}}$.
Then, $D^{(N)} \ket{i,a}^{(N)} = E_i \ket{i,a}^{(N)}$ is approximated by 
the wave equation for an electron in a constant magnetic flux in a flat space:
$ i \left( \gamma^1 (\pa_1+i \sigma^2 B)+\gamma^2 (\pa_2-i \sigma^1 B) \right) \psi_i(\sigma)=E_i \psi_i(\sigma)$,
where $B \sim 1/l_B$ is the flux.
The zero-mode solution (the lowest Landau level) is 
$\psi_0(\sigma) \sim e^{-B z \bar{z}/2} h(z)\left( \begin{array}{c} 0  \\ 1  \end{array} \right)$ 
with $h(z)$ is any function of $z$
and the ``energy'' gap between the excited states and zero-modes will be proportional to $\sqrt{N}$
because of the dimensional analysis.
This state for $h(z)$ is localized at the point if the $h(z)$ is a polynomial of $z$.
Let us consider $\left\langle 0,a \right| f_1(\sigma^\mu) f_2(\sigma^\mu) \ket{0,b}
= (\hat{f}_1 \hat{f}_2)^a_{\,\, b}+\sum_{i \neq 0}  \left\langle 0,a \right| f_1(\sigma^\mu)  \ket{i} \bra{i}   f_2(\sigma^\mu) \ket{0,b} $ where $i$ runs for the excited states.
For the excited states, we see that $\left\langle 0,a \right| f_1(\sigma)  \ket{i}  
=\frac{1}{E_i} \left\langle 0,a \right| [D,f_1(\sigma) ] \ket{i}
= \frac{1}{E_i} \left\langle 0,a \right| \gamma^\mu  {\partial f_1(\sigma) \over \pa \sigma^\mu} \ket{i}$,
thus $\left\langle 0,a \right| f_1(\sigma)  \ket{i}  ={\cal O}(N^{-\frac{1}{2}})$
because ${\partial f_1(\sigma) \over \pa \sigma^\mu} $ does not contains the derivative nor
the gauge field, thus $\left\langle 0,a \right| \gamma^\mu  {\partial f_1(\sigma) \over \pa \sigma^\mu} \ket{i}
={\cal O}(N^0)$.\footnote{
We assumed that the length scale of $f_i(\sigma)$ is much larger than $l_B$. }
This explains (\ref{c1}).

The local physics in a region of the size $l$ with (\ref{size})
is expected to be described by  a flat space approximation with a constant flux.
Here, the number of the zero modes for the flat space is infinite  
because they are parametrized by a function $h(z)$.
On the other hand,
the number of the zero modes for the compact space is generically $N$ by the index theorem, thus finite.
This discrepancy will be manifest if we study some physics at the scale near $l_B$.
Here, because $f_i$ is assumed to be $N$-independent, this approximation will be good.
Let us take a local orthonormal coordinates system $\sigma^\mu$ 
near a point in ${\mathcal M}_{2}$ and
consider $(\hat{\sigma^\mu})^a_{\,\, b} = {}^{(N)} \left\langle 0,a \right| \sigma^\mu \ket{0,b}^{(N)}$ where $\ket{0,a}^{(N)}$ are zero-modes localized near the point on ${\mathcal M}_{2}$.
Then, the flat space with a constant flux approximation gives the canonical commutation relation \cite{Te1}\footnote{
In \cite{Te1}, the momentum $k$ labels the zero-modes which are not localized at a point
and the different gauge was taken.
However, we can use the localized zero-modes, $\ket{n} \sim e^{-B z \bar{z}/2} z^n$, 
where $n$ is non-negative integer, and find the same answer (\ref{coma}).
Indeed, we can see that $\ket{n} $ is orthogonal basis and
$\bra{n} z \ket{m} $ is proportional to $\delta_{m-1,n}$ from 
the phase shift symmetry of $z$. Including the normalization factor, this correctly gives the
commutator of the harmonic oscillator. 
}
\begin{align}
(d A)_{\mu \nu} [\hat{\sigma^\mu}, \hat{\sigma^\nu}] \approx  i \unit.
\label{coma}
\end{align}
The union of the zero-modes which are localized at a point over ${\mathcal M}_{2}$
will span the space of the zero-modes 
and the two zero modes which are localized at different points will have exponentially  
suppressed overwrap.\footnote{
We can check this properties for the fuzzy sphere.
The zero-modes computed in \cite{Te1} for $B>0$ is
$\psi_m \sim \sin(\theta/2)^{(N-1+2m)/2} \cos (\theta/2)^{(N-1-2m)/2} $ 
where $N-1 \geq 2m \geq -(N-1) $.
Because of the second factor, this is localized at $\theta=0$ for  $N-1 \gg 2m \geq -(N-1) $
and the rotational symmetry ensure the localized zero modes near any point.
}
Then, 
the commutator in (\ref{c2}) can be nonzero for the zero modes which are 
close because $f_i(\sigma)$ is a local function.
This contribution is given by
(\ref{coma}) which explains the (\ref{c2}).

\subsection{Matrix regularization and deformation quantization }

We have seen that the the functions on a manifold are regularized to 
matrices by (\ref{W0}) for large, but finite $N$.
In the large $N$ limit, an $1/N$ expansion will be possible.
Let us consider two functions $f_1,f_2$ and 
their matrix regularization $\hat{f}_1=W(f_1),\hat{f}_2=W(f_2)$.
For these, there will be $\tilde{f}_3=f_3^{(0)}+{1 \over N} f_3^{(1)} +{1 \over N^2} f_3^{(2)}+\cdots$
such that $\hat{f}_1 \hat{f}_2 =W(\tilde{f}_3)$
where $f_3^{(i)}$ is $N$-independent function and $f_3^{(0)}=f_1 f_2$.
Then, a non-commutative (associative) product denoted by $*$ for the functions can be defined 
as $f_1 *f_2=\tilde{f}_3$.

On the other hand, the D-branes is describe by the world sheet action
in the perturbative string theory.
In the large flux limit considered in this paper, 
the world sheet action becomes the topological action \cite{Ko, CF, SeWi} which 
gives the deformation quantization of the manifold which give a $*$-product.
Thus, we expect that the our matrix regularization of the geometry
will be equivalent to the deformation quantization in the large $N$ limit.
In particular, we expect that the metric dependence will be negligible in the $1/N$ expansion.  
It would be interesting to study more explicitly the correspondence.

%\section{Conclusion}

\section*{Acknowledgments}

S.T. would like to thank  
for G. Ishiki for 
useful discussions for the contents in section \ref{MG}.
This work was supported by JSPS KAKENHI Grant Number 17K05414.

%\appendix

\section*{Note added:}

As this article was being completed
I was informed that a part of the results in the section 2 
was obtained also in \cite{As}.

\vspace{1cm}

%\noindent
%{\bf Note added}: 

%\appendix

%\section{}

\end{document}